\def\review{0} 
\def\arxivdisclaimer{1} 

\documentclass[10pt, conference, letterpaper]{IEEEtran}
\usepackage[letterpaper, left = 0.68in, right =0.68in, top = 0.7in, bottom = 1in]{geometry}

\usepackage[noadjust]{cite}
\usepackage{amsmath}
\usepackage{amssymb}
\usepackage{amsthm}
\usepackage{amsfonts}
\usepackage{graphicx}
\usepackage{textcomp}
\usepackage{mathtools}
\usepackage{adjustbox}
\usepackage{balance}
\usepackage{bm}
\usepackage{tikz}
\usetikzlibrary{arrows, fit, matrix, positioning, shapes, backgrounds}
\usepackage{xcolor}
\def\BibTeX{{\rm B\kern-.05em{\sc i\kern-.025em b}\kern-.08em
    T\kern-.1667em\lower.7ex\hbox{E}\kern-.125emX}}

\usepackage{textcase}
\usepackage[tablename=TABLE,font=small]{caption}
\DeclareCaptionTextFormat{up}{\MakeTextUppercase{#1}}
\usepackage{paralist}
\usepackage{supertabular}    
\usepackage{array} 
\usepackage[linesnumbered,ruled]{algorithm2e}
\usepackage{algorithmic}
\usepackage[acronym]{glossaries}

\if\review1
\usepackage{todonotes}
\else
\usepackage[disable]{todonotes}
\fi

\if\arxivdisclaimer0
\else
\usepackage{hyperref}
\glsdisablehyper
\fi
\usepackage{cleveref}

\usepackage{numprint}     
\usepackage{makecell}   
\usepackage{colortbl}
\usepackage{enumitem}            
\usepackage{longtable} 
\usepackage{wrapfig}
\usepackage{booktabs}
\usepackage{graphics}
\usepackage{pgfplots}
\usepgfplotslibrary{groupplots}
\usepackage{pgfplotstable}
\usepackage{subcaption}
\pgfplotsset{compat=1.18} 
\usepackage{upgreek}

\usepackage{tabu}

\DeclarePairedDelimiter\ceil{\lceil}{\rceil}

\newtheorem*{remark*}{Remark}

\usepackage[utf8]{inputenc}

\usepackage{balance}
\usepackage{stfloats} 
\usepackage{mathtools}
\usepackage{multirow}

\crefname{figure}{Fig.}{Fig.}
\crefname{table}{Table}{Table}

\let\oldtabular\tabular
\renewcommand{\tabular}{\small\oldtabular}

\newcommand{\egc}{e.\,g., }
\newcommand{\iec}{i.\,e., }

\newcommand{\wrt}{w.\,r.\,t.\ }

\newcommand{\fakepar}[1]{\vspace{1mm}\noindent \textbullet \hspace{1mm}\textit{#1.}}   

\newcommand{\bea}{\begin{eqnarray}} 
\newcommand{\eea}{\end{eqnarray}}

\newcolumntype{?}{!{\vrule width 1pt}}
\definecolor{mittelblau}{RGB}{0, 126, 198}
\definecolor{violettblau}{cmyk}{0.9, 0.6, 0, 0}
\definecolor{rot}{RGB}{238, 28 35}
\definecolor{apfelgruen}{RGB}{140, 198, 62}
\definecolor{gelb}{RGB}{1, 221, 0}
\definecolor{orange}{RGB}{244, 111, 33}
\definecolor{pink}{RGB}{237, 0, 140}
\definecolor{lila}{RGB}{128, 10, 145}
\definecolor{hellgrau}{RGB}{224, 224, 224}
\definecolor{mittelgrau}{RGB}{128, 128, 128}
\definecolor{dunkelgrau}{RGB}{80,80,80}
\definecolor{anthrazit}{RGB}{19, 31, 31}
\definecolor{darkgreen}{RGB}{0.125,0.5,0.169}
\definecolor{ahmedyellow}{RGB}{204,153,0}

\newcommand\blfootnote[1]{%
  \begingroup
  \renewcommand\thefootnote{}\footnote{#1}%
  \addtocounter{footnote}{-1}%
  \endgroup
}

\begin{document}



\pgfplotsset{
    colormap={jet_inue}{
        rgb=(1, 1, 1)
        rgb=(0.99804, 0.99804, 0.99905)
        rgb=(0.99609, 0.99609, 0.99813)
        rgb=(0.99413, 0.99413, 0.99725)
        rgb=(0.99217, 0.99217, 0.99639)
        rgb=(0.99022, 0.99022, 0.99557)
        rgb=(0.98826, 0.98826, 0.99477)
        rgb=(0.9863, 0.9863, 0.99401)
        rgb=(0.98434, 0.98434, 0.99327)
        rgb=(0.98239, 0.98239, 0.99257)
        rgb=(0.98043, 0.98043, 0.9919)
        rgb=(0.97847, 0.97847, 0.99125)
        rgb=(0.97652, 0.97652, 0.99064)
        rgb=(0.97456, 0.97456, 0.99006)
        rgb=(0.9726, 0.9726, 0.98951)
        rgb=(0.97065, 0.97065, 0.98899)
        rgb=(0.96869, 0.96869, 0.9885)
        rgb=(0.96673, 0.96673, 0.98804)
        rgb=(0.96477, 0.96477, 0.98762)
        rgb=(0.96282, 0.96282, 0.98722)
        rgb=(0.96086, 0.96086, 0.98685)
        rgb=(0.9589, 0.9589, 0.98652)
        rgb=(0.95695, 0.95695, 0.98621)
        rgb=(0.95499, 0.95499, 0.98593)
        rgb=(0.95303, 0.95303, 0.98569)
        rgb=(0.95108, 0.95108, 0.98548)
        rgb=(0.94912, 0.94912, 0.98529)
        rgb=(0.94716, 0.94716, 0.98514)
        rgb=(0.94521, 0.94521, 0.98502)
        rgb=(0.94325, 0.94325, 0.98493)
        rgb=(0.94129, 0.94129, 0.98486)
        rgb=(0.93933, 0.93933, 0.98483)
        rgb=(0.93738, 0.93738, 0.98483)
        rgb=(0.93542, 0.93542, 0.98486)
        rgb=(0.93346, 0.93346, 0.98493)
        rgb=(0.93151, 0.93151, 0.98502)
        rgb=(0.92955, 0.92955, 0.98514)
        rgb=(0.92759, 0.92759, 0.98529)
        rgb=(0.92564, 0.92564, 0.98548)
        rgb=(0.92368, 0.92368, 0.98569)
        rgb=(0.92172, 0.92172, 0.98593)
        rgb=(0.91977, 0.91977, 0.98621)
        rgb=(0.91781, 0.91781, 0.98652)
        rgb=(0.91585, 0.91585, 0.98685)
        rgb=(0.91389, 0.91389, 0.98722)
        rgb=(0.91194, 0.91194, 0.98762)
        rgb=(0.90998, 0.90998, 0.98804)
        rgb=(0.90802, 0.90802, 0.9885)
        rgb=(0.90607, 0.90607, 0.98899)
        rgb=(0.90411, 0.90411, 0.98951)
        rgb=(0.90215, 0.90215, 0.99006)
        rgb=(0.9002, 0.9002, 0.99064)
        rgb=(0.89824, 0.89824, 0.99125)
        rgb=(0.89628, 0.89628, 0.9919)
        rgb=(0.89432, 0.89432, 0.99257)
        rgb=(0.89237, 0.89237, 0.99327)
        rgb=(0.89041, 0.89041, 0.99401)
        rgb=(0.88845, 0.88845, 0.99477)
        rgb=(0.8865, 0.8865, 0.99557)
        rgb=(0.88454, 0.88454, 0.99639)
        rgb=(0.88258, 0.88258, 0.99725)
        rgb=(0.88063, 0.88063, 0.99813)
        rgb=(0.87867, 0.87867, 0.99905)
        rgb=(0.87671, 0.87671, 1)
        rgb=(0.87476, 0.87573, 1)
        rgb=(0.8728, 0.87479, 1)
        rgb=(0.87084, 0.87387, 1)
        rgb=(0.86888, 0.87298, 1)
        rgb=(0.86693, 0.87213, 1)
        rgb=(0.86497, 0.8713, 1)
        rgb=(0.86301, 0.87051, 1)
        rgb=(0.86106, 0.86974, 1)
        rgb=(0.8591, 0.86901, 1)
        rgb=(0.85714, 0.8683, 1)
        rgb=(0.85519, 0.86763, 1)
        rgb=(0.85323, 0.86699, 1)
        rgb=(0.85127, 0.86638, 1)
        rgb=(0.84932, 0.8658, 1)
        rgb=(0.84736, 0.86525, 1)
        rgb=(0.8454, 0.86473, 1)
        rgb=(0.84344, 0.86424, 1)
        rgb=(0.84149, 0.86378, 1)
        rgb=(0.83953, 0.86335, 1)
        rgb=(0.83757, 0.86295, 1)
        rgb=(0.83562, 0.86259, 1)
        rgb=(0.83366, 0.86225, 1)
        rgb=(0.8317, 0.86194, 1)
        rgb=(0.82975, 0.86167, 1)
        rgb=(0.82779, 0.86142, 1)
        rgb=(0.82583, 0.86121, 1)
        rgb=(0.82387, 0.86103, 1)
        rgb=(0.82192, 0.86087, 1)
        rgb=(0.81996, 0.86075, 1)
        rgb=(0.818, 0.86066, 1)
        rgb=(0.81605, 0.8606, 1)
        rgb=(0.81409, 0.86057, 1)
        rgb=(0.81213, 0.86057, 1)
        rgb=(0.81018, 0.8606, 1)
        rgb=(0.80822, 0.86066, 1)
        rgb=(0.80626, 0.86075, 1)
        rgb=(0.80431, 0.86087, 1)
        rgb=(0.80235, 0.86103, 1)
        rgb=(0.80039, 0.86121, 1)
        rgb=(0.79843, 0.86142, 1)
        rgb=(0.79648, 0.86167, 1)
        rgb=(0.79452, 0.86194, 1)
        rgb=(0.79256, 0.86225, 1)
        rgb=(0.79061, 0.86259, 1)
        rgb=(0.78865, 0.86295, 1)
        rgb=(0.78669, 0.86335, 1)
        rgb=(0.78474, 0.86378, 1)
        rgb=(0.78278, 0.86424, 1)
        rgb=(0.78082, 0.86473, 1)
        rgb=(0.77886, 0.86525, 1)
        rgb=(0.77691, 0.8658, 1)
        rgb=(0.77495, 0.86638, 1)
        rgb=(0.77299, 0.86699, 1)
        rgb=(0.77104, 0.86763, 1)
        rgb=(0.76908, 0.8683, 1)
        rgb=(0.76712, 0.86901, 1)
        rgb=(0.76517, 0.86974, 1)
        rgb=(0.76321, 0.87051, 1)
        rgb=(0.76125, 0.8713, 1)
        rgb=(0.7593, 0.87213, 1)
        rgb=(0.75734, 0.87298, 1)
        rgb=(0.75538, 0.87387, 1)
        rgb=(0.75342, 0.87479, 1)
        rgb=(0.75147, 0.87573, 1)
        rgb=(0.74951, 0.87671, 1)
        rgb=(0.74755, 0.87772, 1)
        rgb=(0.7456, 0.87876, 1)
        rgb=(0.74364, 0.87983, 1)
        rgb=(0.74168, 0.88093, 1)
        rgb=(0.73973, 0.88206, 1)
        rgb=(0.73777, 0.88323, 1)
        rgb=(0.73581, 0.88442, 1)
        rgb=(0.73386, 0.88564, 1)
        rgb=(0.7319, 0.88689, 1)
        rgb=(0.72994, 0.88818, 1)
        rgb=(0.72798, 0.88949, 1)
        rgb=(0.72603, 0.89084, 1)
        rgb=(0.72407, 0.89222, 1)
        rgb=(0.72211, 0.89362, 1)
        rgb=(0.72016, 0.89506, 1)
        rgb=(0.7182, 0.89653, 1)
        rgb=(0.71624, 0.89802, 1)
        rgb=(0.71429, 0.89955, 1)
        rgb=(0.71233, 0.90111, 1)
        rgb=(0.71037, 0.9027, 1)
        rgb=(0.70841, 0.90432, 1)
        rgb=(0.70646, 0.90597, 1)
        rgb=(0.7045, 0.90766, 1)
        rgb=(0.70254, 0.90937, 1)
        rgb=(0.70059, 0.91111, 1)
        rgb=(0.69863, 0.91289, 1)
        rgb=(0.69667, 0.91469, 1)
        rgb=(0.69472, 0.91652, 1)
        rgb=(0.69276, 0.91839, 1)
        rgb=(0.6908, 0.92028, 1)
        rgb=(0.68885, 0.92221, 1)
        rgb=(0.68689, 0.92417, 1)
        rgb=(0.68493, 0.92616, 1)
        rgb=(0.68297, 0.92817, 1)
        rgb=(0.68102, 0.93022, 1)
        rgb=(0.67906, 0.9323, 1)
        rgb=(0.6771, 0.93441, 1)
        rgb=(0.67515, 0.93655, 1)
        rgb=(0.67319, 0.93872, 1)
        rgb=(0.67123, 0.94092, 1)
        rgb=(0.66928, 0.94316, 1)
        rgb=(0.66732, 0.94542, 1)
        rgb=(0.66536, 0.94771, 1)
        rgb=(0.66341, 0.95004, 1)
        rgb=(0.66145, 0.95239, 1)
        rgb=(0.65949, 0.95478, 1)
        rgb=(0.65753, 0.95719, 1)
        rgb=(0.65558, 0.95964, 1)
        rgb=(0.65362, 0.96211, 1)
        rgb=(0.65166, 0.96462, 1)
        rgb=(0.64971, 0.96716, 1)
        rgb=(0.64775, 0.96973, 1)
        rgb=(0.64579, 0.97233, 1)
        rgb=(0.64384, 0.97496, 1)
        rgb=(0.64188, 0.97762, 1)
        rgb=(0.63992, 0.98031, 1)
        rgb=(0.63796, 0.98303, 1)
        rgb=(0.63601, 0.98578, 1)
        rgb=(0.63405, 0.98856, 1)
        rgb=(0.63209, 0.99138, 1)
        rgb=(0.63014, 0.99422, 1)
        rgb=(0.62818, 0.9971, 1)
        rgb=(0.62622, 1, 1)
        rgb=(0.6272, 1, 0.99706)
        rgb=(0.62821, 1, 0.9941)
        rgb=(0.62925, 1, 0.9911)
        rgb=(0.63032, 1, 0.98807)
        rgb=(0.63142, 1, 0.98502)
        rgb=(0.63255, 1, 0.98193)
        rgb=(0.63371, 1, 0.97881)
        rgb=(0.63491, 1, 0.97566)
        rgb=(0.63613, 1, 0.97248)
        rgb=(0.63738, 1, 0.96927)
        rgb=(0.63867, 1, 0.96603)
        rgb=(0.63998, 1, 0.96276)
        rgb=(0.64133, 1, 0.95945)
        rgb=(0.6427, 1, 0.95612)
        rgb=(0.64411, 1, 0.95276)
        rgb=(0.64555, 1, 0.94936)
        rgb=(0.64702, 1, 0.94594)
        rgb=(0.64851, 1, 0.94248)
        rgb=(0.65004, 1, 0.939)
        rgb=(0.6516, 1, 0.93548)
        rgb=(0.65319, 1, 0.93193)
        rgb=(0.65481, 1, 0.92836)
        rgb=(0.65646, 1, 0.92475)
        rgb=(0.65815, 1, 0.92111)
        rgb=(0.65986, 1, 0.91744)
        rgb=(0.6616, 1, 0.91374)
        rgb=(0.66337, 1, 0.91001)
        rgb=(0.66518, 1, 0.90625)
        rgb=(0.66701, 1, 0.90246)
        rgb=(0.66888, 1, 0.89864)
        rgb=(0.67077, 1, 0.89478)
        rgb=(0.6727, 1, 0.8909)
        rgb=(0.67466, 1, 0.88699)
        rgb=(0.67665, 1, 0.88304)
        rgb=(0.67866, 1, 0.87907)
        rgb=(0.68071, 1, 0.87506)
        rgb=(0.68279, 1, 0.87102)
        rgb=(0.6849, 1, 0.86696)
        rgb=(0.68704, 1, 0.86286)
        rgb=(0.68921, 1, 0.85873)
        rgb=(0.69141, 1, 0.85457)
        rgb=(0.69365, 1, 0.85039)
        rgb=(0.69591, 1, 0.84617)
        rgb=(0.6982, 1, 0.84192)
        rgb=(0.70053, 1, 0.83763)
        rgb=(0.70288, 1, 0.83332)
        rgb=(0.70527, 1, 0.82898)
        rgb=(0.70768, 1, 0.82461)
        rgb=(0.71013, 1, 0.82021)
        rgb=(0.7126, 1, 0.81577)
        rgb=(0.71511, 1, 0.81131)
        rgb=(0.71765, 1, 0.80681)
        rgb=(0.72022, 1, 0.80229)
        rgb=(0.72282, 1, 0.79773)
        rgb=(0.72545, 1, 0.79314)
        rgb=(0.72811, 1, 0.78853)
        rgb=(0.7308, 1, 0.78388)
        rgb=(0.73352, 1, 0.7792)
        rgb=(0.73627, 1, 0.77449)
        rgb=(0.73905, 1, 0.76975)
        rgb=(0.74187, 1, 0.76498)
        rgb=(0.74471, 1, 0.76018)
        rgb=(0.74758, 1, 0.75535)
        rgb=(0.75049, 1, 0.75049)
        rgb=(0.75342, 1, 0.7456)
        rgb=(0.75639, 1, 0.74067)
        rgb=(0.75939, 1, 0.73572)
        rgb=(0.76241, 1, 0.73074)
        rgb=(0.76547, 1, 0.72572)
        rgb=(0.76856, 1, 0.72068)
        rgb=(0.77168, 1, 0.7156)
        rgb=(0.77483, 1, 0.71049)
        rgb=(0.77801, 1, 0.70536)
        rgb=(0.78122, 1, 0.70019)
        rgb=(0.78446, 1, 0.69499)
        rgb=(0.78773, 1, 0.68976)
        rgb=(0.79103, 1, 0.6845)
        rgb=(0.79437, 1, 0.67921)
        rgb=(0.79773, 1, 0.67389)
        rgb=(0.80113, 1, 0.66854)
        rgb=(0.80455, 1, 0.66316)
        rgb=(0.80801, 1, 0.65775)
        rgb=(0.81149, 1, 0.65231)
        rgb=(0.81501, 1, 0.64683)
        rgb=(0.81855, 1, 0.64133)
        rgb=(0.82213, 1, 0.63579)
        rgb=(0.82574, 1, 0.63023)
        rgb=(0.82938, 1, 0.62463)
        rgb=(0.83305, 1, 0.61901)
        rgb=(0.83675, 1, 0.61335)
        rgb=(0.84048, 1, 0.60766)
        rgb=(0.84424, 1, 0.60194)
        rgb=(0.84803, 1, 0.5962)
        rgb=(0.85185, 1, 0.59042)
        rgb=(0.85571, 1, 0.58461)
        rgb=(0.85959, 1, 0.57877)
        rgb=(0.8635, 1, 0.5729)
        rgb=(0.86745, 1, 0.56699)
        rgb=(0.87142, 1, 0.56106)
        rgb=(0.87543, 1, 0.5551)
        rgb=(0.87946, 1, 0.54911)
        rgb=(0.88353, 1, 0.54308)
        rgb=(0.88763, 1, 0.53703)
        rgb=(0.89176, 1, 0.53094)
        rgb=(0.89591, 1, 0.52483)
        rgb=(0.9001, 1, 0.51868)
        rgb=(0.90432, 1, 0.51251)
        rgb=(0.90857, 1, 0.5063)
        rgb=(0.91285, 1, 0.50006)
        rgb=(0.91717, 1, 0.49379)
        rgb=(0.92151, 1, 0.48749)
        rgb=(0.92588, 1, 0.48116)
        rgb=(0.93028, 1, 0.4748)
        rgb=(0.93472, 1, 0.46841)
        rgb=(0.93918, 1, 0.46199)
        rgb=(0.94368, 1, 0.45554)
        rgb=(0.9482, 1, 0.44906)
        rgb=(0.95276, 1, 0.44255)
        rgb=(0.95734, 1, 0.436)
        rgb=(0.96196, 1, 0.42943)
        rgb=(0.96661, 1, 0.42282)
        rgb=(0.97129, 1, 0.41619)
        rgb=(0.976, 1, 0.40952)
        rgb=(0.98074, 1, 0.40283)
        rgb=(0.98551, 1, 0.3961)
        rgb=(0.99031, 1, 0.38934)
        rgb=(0.99514, 1, 0.38255)
        rgb=(1, 1, 0.37573)
        rgb=(1, 0.99511, 0.37378)
        rgb=(1, 0.99018, 0.37182)
        rgb=(1, 0.98523, 0.36986)
        rgb=(1, 0.98025, 0.36791)
        rgb=(1, 0.97523, 0.36595)
        rgb=(1, 0.97019, 0.36399)
        rgb=(1, 0.96511, 0.36204)
        rgb=(1, 0.96, 0.36008)
        rgb=(1, 0.95487, 0.35812)
        rgb=(1, 0.9497, 0.35616)
        rgb=(1, 0.9445, 0.35421)
        rgb=(1, 0.93927, 0.35225)
        rgb=(1, 0.93401, 0.35029)
        rgb=(1, 0.92872, 0.34834)
        rgb=(1, 0.9234, 0.34638)
        rgb=(1, 0.91805, 0.34442)
        rgb=(1, 0.91267, 0.34247)
        rgb=(1, 0.90726, 0.34051)
        rgb=(1, 0.90182, 0.33855)
        rgb=(1, 0.89634, 0.33659)
        rgb=(1, 0.89084, 0.33464)
        rgb=(1, 0.8853, 0.33268)
        rgb=(1, 0.87974, 0.33072)
        rgb=(1, 0.87414, 0.32877)
        rgb=(1, 0.86852, 0.32681)
        rgb=(1, 0.86286, 0.32485)
        rgb=(1, 0.85717, 0.3229)
        rgb=(1, 0.85146, 0.32094)
        rgb=(1, 0.84571, 0.31898)
        rgb=(1, 0.83993, 0.31703)
        rgb=(1, 0.83412, 0.31507)
        rgb=(1, 0.82828, 0.31311)
        rgb=(1, 0.82241, 0.31115)
        rgb=(1, 0.81651, 0.3092)
        rgb=(1, 0.81057, 0.30724)
        rgb=(1, 0.80461, 0.30528)
        rgb=(1, 0.79862, 0.30333)
        rgb=(1, 0.79259, 0.30137)
        rgb=(1, 0.78654, 0.29941)
        rgb=(1, 0.78045, 0.29746)
        rgb=(1, 0.77434, 0.2955)
        rgb=(1, 0.76819, 0.29354)
        rgb=(1, 0.76202, 0.29159)
        rgb=(1, 0.75581, 0.28963)
        rgb=(1, 0.74957, 0.28767)
        rgb=(1, 0.7433, 0.28571)
        rgb=(1, 0.737, 0.28376)
        rgb=(1, 0.73068, 0.2818)
        rgb=(1, 0.72432, 0.27984)
        rgb=(1, 0.71792, 0.27789)
        rgb=(1, 0.7115, 0.27593)
        rgb=(1, 0.70505, 0.27397)
        rgb=(1, 0.69857, 0.27202)
        rgb=(1, 0.69206, 0.27006)
        rgb=(1, 0.68551, 0.2681)
        rgb=(1, 0.67894, 0.26614)
        rgb=(1, 0.67233, 0.26419)
        rgb=(1, 0.6657, 0.26223)
        rgb=(1, 0.65903, 0.26027)
        rgb=(1, 0.65234, 0.25832)
        rgb=(1, 0.64561, 0.25636)
        rgb=(1, 0.63885, 0.2544)
        rgb=(1, 0.63206, 0.25245)
        rgb=(1, 0.62524, 0.25049)
        rgb=(1, 0.6184, 0.24853)
        rgb=(1, 0.61152, 0.24658)
        rgb=(1, 0.6046, 0.24462)
        rgb=(1, 0.59766, 0.24266)
        rgb=(1, 0.59069, 0.2407)
        rgb=(1, 0.58369, 0.23875)
        rgb=(1, 0.57666, 0.23679)
        rgb=(1, 0.56959, 0.23483)
        rgb=(1, 0.5625, 0.23288)
        rgb=(1, 0.55538, 0.23092)
        rgb=(1, 0.54822, 0.22896)
        rgb=(1, 0.54103, 0.22701)
        rgb=(1, 0.53382, 0.22505)
        rgb=(1, 0.52657, 0.22309)
        rgb=(1, 0.51929, 0.22114)
        rgb=(1, 0.51199, 0.21918)
        rgb=(1, 0.50465, 0.21722)
        rgb=(1, 0.49728, 0.21526)
        rgb=(1, 0.48988, 0.21331)
        rgb=(1, 0.48245, 0.21135)
        rgb=(1, 0.47499, 0.20939)
        rgb=(1, 0.4675, 0.20744)
        rgb=(1, 0.45997, 0.20548)
        rgb=(1, 0.45242, 0.20352)
        rgb=(1, 0.44484, 0.20157)
        rgb=(1, 0.43722, 0.19961)
        rgb=(1, 0.42958, 0.19765)
        rgb=(1, 0.42191, 0.19569)
        rgb=(1, 0.4142, 0.19374)
        rgb=(1, 0.40646, 0.19178)
        rgb=(1, 0.3987, 0.18982)
        rgb=(1, 0.3909, 0.18787)
        rgb=(1, 0.38307, 0.18591)
        rgb=(1, 0.37521, 0.18395)
        rgb=(1, 0.36733, 0.182)
        rgb=(1, 0.35941, 0.18004)
        rgb=(1, 0.35146, 0.17808)
        rgb=(1, 0.34347, 0.17613)
        rgb=(1, 0.33546, 0.17417)
        rgb=(1, 0.32742, 0.17221)
        rgb=(1, 0.31935, 0.17025)
        rgb=(1, 0.31125, 0.1683)
        rgb=(1, 0.30311, 0.16634)
        rgb=(1, 0.29495, 0.16438)
        rgb=(1, 0.28675, 0.16243)
        rgb=(1, 0.27853, 0.16047)
        rgb=(1, 0.27027, 0.15851)
        rgb=(1, 0.26199, 0.15656)
        rgb=(1, 0.25367, 0.1546)
        rgb=(1, 0.24532, 0.15264)
        rgb=(1, 0.23694, 0.15068)
        rgb=(1, 0.22853, 0.14873)
        rgb=(1, 0.2201, 0.14677)
        rgb=(1, 0.21163, 0.14481)
        rgb=(1, 0.20312, 0.14286)
        rgb=(1, 0.19459, 0.1409)
        rgb=(1, 0.18603, 0.13894)
        rgb=(1, 0.17744, 0.13699)
        rgb=(1, 0.16882, 0.13503)
        rgb=(1, 0.16016, 0.13307)
        rgb=(1, 0.15148, 0.13112)
        rgb=(1, 0.14277, 0.12916)
        rgb=(1, 0.13402, 0.1272)
        rgb=(1, 0.12524, 0.12524)
        rgb=(0.99315, 0.12329, 0.12329)
        rgb=(0.98627, 0.12133, 0.12133)
        rgb=(0.97936, 0.11937, 0.11937)
        rgb=(0.97242, 0.11742, 0.11742)
        rgb=(0.96545, 0.11546, 0.11546)
        rgb=(0.95845, 0.1135, 0.1135)
        rgb=(0.95141, 0.11155, 0.11155)
        rgb=(0.94435, 0.10959, 0.10959)
        rgb=(0.93726, 0.10763, 0.10763)
        rgb=(0.93013, 0.10568, 0.10568)
        rgb=(0.92298, 0.10372, 0.10372)
        rgb=(0.91579, 0.10176, 0.10176)
        rgb=(0.90857, 0.099804, 0.099804)
        rgb=(0.90133, 0.097847, 0.097847)
        rgb=(0.89405, 0.09589, 0.09589)
        rgb=(0.88674, 0.093933, 0.093933)
        rgb=(0.8794, 0.091977, 0.091977)
        rgb=(0.87203, 0.09002, 0.09002)
        rgb=(0.86463, 0.088063, 0.088063)
        rgb=(0.8572, 0.086106, 0.086106)
        rgb=(0.84974, 0.084149, 0.084149)
        rgb=(0.84225, 0.082192, 0.082192)
        rgb=(0.83473, 0.080235, 0.080235)
        rgb=(0.82718, 0.078278, 0.078278)
        rgb=(0.81959, 0.076321, 0.076321)
        rgb=(0.81198, 0.074364, 0.074364)
        rgb=(0.80434, 0.072407, 0.072407)
        rgb=(0.79666, 0.07045, 0.07045)
        rgb=(0.78896, 0.068493, 0.068493)
        rgb=(0.78122, 0.066536, 0.066536)
        rgb=(0.77345, 0.064579, 0.064579)
        rgb=(0.76566, 0.062622, 0.062622)
        rgb=(0.75783, 0.060665, 0.060665)
        rgb=(0.74997, 0.058708, 0.058708)
        rgb=(0.74208, 0.056751, 0.056751)
        rgb=(0.73416, 0.054795, 0.054795)
        rgb=(0.72621, 0.052838, 0.052838)
        rgb=(0.71823, 0.050881, 0.050881)
        rgb=(0.71022, 0.048924, 0.048924)
        rgb=(0.70218, 0.046967, 0.046967)
        rgb=(0.6941, 0.04501, 0.04501)
        rgb=(0.686, 0.043053, 0.043053)
        rgb=(0.67787, 0.041096, 0.041096)
        rgb=(0.6697, 0.039139, 0.039139)
        rgb=(0.66151, 0.037182, 0.037182)
        rgb=(0.65328, 0.035225, 0.035225)
        rgb=(0.64503, 0.033268, 0.033268)
        rgb=(0.63674, 0.031311, 0.031311)
        rgb=(0.62842, 0.029354, 0.029354)
        rgb=(0.62008, 0.027397, 0.027397)
        rgb=(0.6117, 0.02544, 0.02544)
        rgb=(0.60329, 0.023483, 0.023483)
        rgb=(0.59485, 0.021526, 0.021526)
        rgb=(0.58638, 0.019569, 0.019569)
        rgb=(0.57788, 0.017613, 0.017613)
        rgb=(0.56935, 0.015656, 0.015656)
        rgb=(0.56079, 0.013699, 0.013699)
        rgb=(0.5522, 0.011742, 0.011742)
        rgb=(0.54357, 0.0097847, 0.0097847)
        rgb=(0.53492, 0.0078278, 0.0078278)
        rgb=(0.52624, 0.0058708, 0.0058708)
        rgb=(0.51752, 0.0039139, 0.0039139)
        rgb=(0.50878, 0.0019569, 0.0019569)
        rgb=(0.5, 0, 0)
    }
}

\title{Feasibility of Non-Line-of-Sight Integrated Sensing and
Communication at mmWave}

\author{
    \IEEEauthorblockN{
        Paolo Tosi\IEEEauthorrefmark{1}\IEEEauthorrefmark{2},
        Marcus Henninger\IEEEauthorrefmark{1},
        Lucas Giroto de Oliveira\IEEEauthorrefmark{3},
        Silvio Mandelli\IEEEauthorrefmark{1}
        }
	\IEEEauthorblockA{
    \IEEEauthorrefmark{1}Nokia Bell Labs Stuttgart, Germany 
\;\;
    \IEEEauthorrefmark{2}Politecnico di Milano, Italy
    \\
    \IEEEauthorrefmark{3}Karlsruhe Institute of Technology (KIT), Germany \\
	E-mail: paolo.tosi@nokia.com, marcus.henninger@nokia.com}
 }

\maketitle



\newacronym{3GPP}{3GPP}{3rd Generation Partnership Project}
\newacronym{5G}{5G}{fifth generation}
\newacronym{6G}{6G}{sixth generation}
\newacronym{awgn}{AWGN}{additive white Gaussian noise}
\newacronym{bts}{BTS}{base transceiver station}
\newacronym{cfar}{CFAR}{constant false alarm rate}
\newacronym{cp}{CP}{cyclic prefix}
\newacronym{csi}{CSI}{channel state information}
\newacronym{dl}{DL}{downlink}
\newacronym{dft}{DFT}{Discrete Fourier Transform}
\newacronym{fmcw}{FMCW}{frequency-modulated continuous-wave}
\newacronym{fr2}{FR2}{frequency range~2}
\newacronym{idft}{IDFT}{inverse discrete Fourier transform}
\newacronym{gnb}{gNB}{gNodeB}
\newacronym{isac}{ISAC}{Integrated Sensing and Communication}
\newacronym{los}{LOS}{line-of-sight}
\newacronym{mmw}{mmWave}{millimeter wave}
\newacronym{nlos}{NLOS}{non-line-of-sight}
\newacronym{ofdm}{OFDM}{orthogonal frequency-division multiplexing}
\newacronym{poc}{POC}{proof of concept}
\newacronym{ru}{RU}{radio unit}
\newacronym{rx}{RX}{receiver}
\newacronym{snr}{SNR}{signal-to-noise ratio}
\newacronym{spu}{SPU}{sensing processing unit}
\newacronym{tdd}{TDD}{time division duplex}
\newacronym{tx}{TX}{transmitter}
\newacronym{ue}{UE}{user equipment}
\newacronym{ul}{UL}{uplink}

\begin{abstract}

One rarely addressed direction in the context of \gls{isac} is \gls{nlos} sensing, with the potential to enable use cases like intrusion detection and to increase the value that wireless networks can bring. However, \gls{isac} networks impose challenges for sensing due to their communication-oriented design. For instance, time division duplex transmission creates spectral holes in time, resulting in spectral replicas in the radar image. To counteract this, we evaluate different channel state information processing strategies and discuss their tradeoffs. We further propose an ensemble of techniques to detect targets in \gls{nlos} conditions. Our approaches are validated with experiments using a millimeter wave \gls{isac} proof of concept in a factory-like environment. The results show that target detection in \gls{nlos} is generally possible with \gls{isac}.


\end{abstract}



\if\arxivdisclaimer1
\blfootnote{© 20XX IEEE.  Personal use of this material is permitted.  Permission from IEEE must be obtained for all other uses, in any current or future media, including reprinting/republishing this material for advertising or promotional purposes, creating new collective works, for resale or redistribution to servers or lists, or reuse of any copyrighted component of this work in other works.}
\else
\vspace{0.2cm}
\fi


\glsresetall

\section{Introduction}\label{sec:introduction}

One of the new features of upcoming \gls{6G} \gls{3GPP} cellular networks is the ability to gather information about the environment, essentially operating the network as a radar.
The integration of this functionality into the existing cellular communications infrastructure is referred to as \gls{isac}.
Recent research has been focusing, among other aspects, on the definition of plausible use cases for \gls{isac} systems and their requirements~\cite{Mandelli_Henninger_Bauhofer_Wild_2023, Wang_isac_survey_2022}. 

The typical assumption for most \gls{isac} use cases is that targets are in \gls{los} from \gls{tx} and \gls{rx}.
However, radio signals may reach targets only indirectly via (multiple) reflections, \egc off walls or buildings. Even though this would allow to acquire some information about the targets, \gls{nlos} sensing in \gls{isac} is still a rather unexplored area of research.
The main opportunity offered by \gls{nlos} sensing is the ability to enable use cases that are not possible with fixed camera or LiDAR systems, like ``around-the-corner'' detection. 
This feature would be particularly useful in applications such as safe urban mobility and surveillance systems for intrusion detection, where it is necessary to detect targets moving behind obstacles. In addition, the ability to operate in \gls{nlos} conditions would allow the reuse of the same infrastructure deployed for communication purposes without the need for excessive densification.

Different \gls{nlos} sensing approaches using radar radios have been proposed in literature.
For instance, target features such as breathing rate were extracted in \gls{nlos} conditions in~\cite{Li_Ge_Wang_BreathingNLOS_2022} using frequency-modulated continuous-wave radar operating in \gls{mmw}. 
In~\cite{Gustafsson_Doppler_urban_2016}, micro-Doppler signatures of multiple human targets were extracted in a controlled urban scenario with an experimental coherent high-resolution X-band radar.
A solution for performing \gls{nlos} detection of vehicles using a mmWave \gls{ofdm} radar was presented in \cite{Solomitckii_nlos_uwb}, exploiting planar reflectors positioned at an L-shaped intersection. In a recent study, the authors of~\cite{pegoraro2024jump} used an experimental 60~GHz platform for human tracking in \gls{nlos} conditions indoors.

While the previous studies have shown promising results, they were performed using dedicated radar radios. 
However, the first envisioned cellular \gls{isac} systems present substantially different challenges, since they are designed for communications and not yet optimized for sensing. For instance, they are bound to legacy frame structures and numerologies originally designed for communications. An example of this is the use of \gls{tdd}, which creates spectral holes in time and requires dedicated \gls{csi} processing techniques to avoid undesired spectral replicas. 
This issue raises the distortion level in the radar image, impairing the feasibility of \gls{snr}-limited use cases, like \gls{nlos} sensing or drone detection.


To overcome this, we utilize \gls{csi} decimation to avoid spectral holes in time, thereby removing the replicas. We investigate the tradeoffs of different \gls{csi} processing techniques \wrt \gls{snr} and achievable sensing performance. Using different radar processing techniques to enable target detection, we leverage the resulting \gls{snr} and resolution gain to demonstrate the feasibility of intrusion detection based on \gls{isac} in \gls{nlos} conditions. The experiments conducted with a \gls{mmw} \gls{isac} \gls{poc} in an indoor environment show that target detection in \gls{nlos} is possible and indicate the fundamental feasibility of intrusion detection in \gls{isac}.




\section{ISAC Setup}\label{sec:PoCScenario}

This work was developed using measurements from our \gls{isac} \gls{poc}~\cite{wild2023integrated}, based on commercially
available \gls{5G} communications hardware in \gls{fr2} at central frequency $f_c = 27.4$~GHz.
The system comprises a half-duplex \gls{gnb} \gls{ru} acting as \gls{tx}, extended by a sniffer \gls{ru} serving as \gls{rx}.
The \glspl{ru} are physically separated, but quasi co-located, allowing to treat the system as a mono-static sensing setup. The \gls{gnb} \gls{ru} transmits \gls{5G} compliant \gls{ofdm} radio frames with $T_f = 10$~ms duration in \gls{tdd}. Each \gls{tdd} pattern extends over $T_{\text{TDD}} = 1.25$~ms~\cite{3gpp_38331} and comprises $M_{\text{DL}} = 104$ \gls{dl} and $M_{\text{UL}} = 36$ \gls{ul} symbols, \iec a \gls{dl}/\gls{ul} ratio of ca. 3:1.
The sniffer \gls{ru} is synchronized with the \gls{gnb} \gls{ru} and it always operates in \gls{ul}. 
A dedicated server receives the transmitted and received signals as complex IQ samples from \gls{gnb} \gls{ru} and sniffer \gls{ru}, respectively, and computes the \gls{csi} matrix $\mathbf H \in \mathbb{C}^{N\times M}$ via division of the reflected signal by the transmitted signal per frame. Moreover, \gls{gnb} \gls{ru} and sniffer \gls{ru} use the same fixed beam (with 14\textdegree~half-power horizontal beam width) for transmitting the signal and receiving the reflections, respectively.
The \gls{poc} parameters based on numerology $\mu = 3$~\cite{3gpp_38211} are listed in Tab.~~\ref{table:PoC_params}. For more details, please refer to~\cite{wild2023integrated}. 

\begin{table}[b]
    \caption{POC System parameters.}\label{table:PoC_params}
	\centering
     \begin{tabu}{|l|l|r|}
            \hline
            \textbf{Parameter} & \textbf{Description} & \textbf{Value} \\
			\Xhline{3\arrayrulewidth}
			$f_c$ & Carrier frequency & 27.4 GHz \\
  			\hline
            $B$ & Total bandwidth $N\cdot \Delta f$ & 190 MHz \\
  			\hline
  			$N$ & Number of subcarriers & 1584 \\
            \hline
  			$M$ & Number of \gls{ofdm} symbols per radio frame & 1120\\
  			\hline
  			$\Delta f$ & Subcarrier spacing & 120 kHz\\
  			\hline
  			$T_0$ & OFDM symbol time & 8.33 $\upmu$s\\
            \hline
            $T_{\text{CP}}$ & CP length & 0.59 $\upmu$s \\
  			\hline
  			$T_{s}$ & OFDM symbol time including CP & 8.92 $\upmu$s \\
  			\hline
    \end{tabu}
\end{table}


\begin{figure}[t]
    \setlength{\abovecaptionskip}{3pt} 
    \setlength{\belowcaptionskip}{3pt} 
    \centering
    \includegraphics[width=0.8\columnwidth]{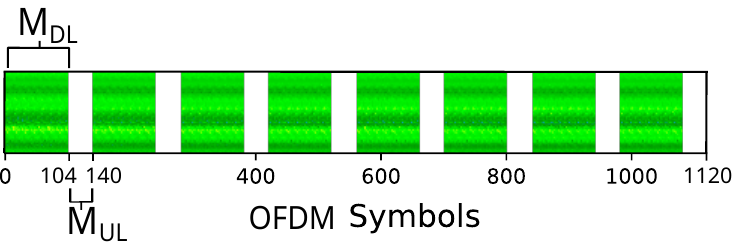}
    \caption{\small \gls{csi} visualization of \gls{tdd} pattern of one frame. Green columns represent \acrshort{dl} symbols, blank columns correspond to UL symbols.}
    \label{fig:CSI_with_TDD_patt}
\end{figure}


The achievable system performance can be evaluated in terms of resolution and unambiguous aperture for both range and speed.
Unambiguous range $d_{\text{unamb}}$ and speed $v_{\text{unamb}}$ are the maximum values without ambiguities due to aliasing, and depend on the sampling rate in frequency and time domain, defined by the subcarrier spacing $\Delta f$ and the \gls{ofdm} symbol time $T_s$, respectively. Range and speed resolution, $d_{\text{res}}$ and $v_{\text{res}}$, refer to the ability to discriminate targets based on their range and speed. The resolution depends on the aperture in frequency and time, given by the bandwidth $N\Delta f$ and observation time $M T_S$, respectively~\cite{de2021joint}. 
Computing the periodogram~\cite{braun2014ofdm} by processing the \gls{csi} of a single frame with the parameters from Tab.~\ref{table:PoC_params},
the \gls{poc} operates with the following performance
\begin{align}
v_{\text{unamb}} &= \frac{c_0}{2f_C T_s} = 613.5~\text{m/s} \; , \label{eq:speed_unamb} \\
d_{\text{unamb}} &= \frac{c_0}{2\Delta f} = 1250~\text{m} \; , \\
v_{\text{res}} &= \frac{c_0}{2MT_sf_C} = 0.55 \text{ m/s} \; , \label{eq:speed_res} \\
d_{\text{res}} &= \frac{c_0}{2N\Delta f} = 0.79~\text{m} \; .
\end{align}
Another important parameter is the \gls{snr} in the periodogram, which we define as the ratio between the strongest return in the periodogram and the measured noise level in the periodogram
\begin{align}
    \text{SNR} = \frac{\max{}(S(n,m))}{\sigma_N^2} \; .
    \label{eq:snr}
\end{align}
The periodogram \gls{snr} includes a multiplicative processing gain given by the number of subcarriers and \gls{ofdm} symbols. Thus, to maximize this gain, one would like to use the full \gls{csi} matrix $\mathbf{H}$ for sensing.
A high \gls{snr} is is essential to distinguish targets from noise and especially critical for \gls{nlos} sensing, where the received reflected signal power is typically low due to the propagation distance and multiple reflections.

\section{Time Division Duplex in ISAC}\label{sec:csi_strategies}

\subsection{Impact of Time Domain Holes}\label{sec:tdd_impact}

\begin{figure}[t]
    \setlength{\abovecaptionskip}{3pt} 
    \setlength{\belowcaptionskip}{3pt} 
    \setlength{\intextsep}{4pt} 
    
    \centering
    \includegraphics[width=0.82\columnwidth]{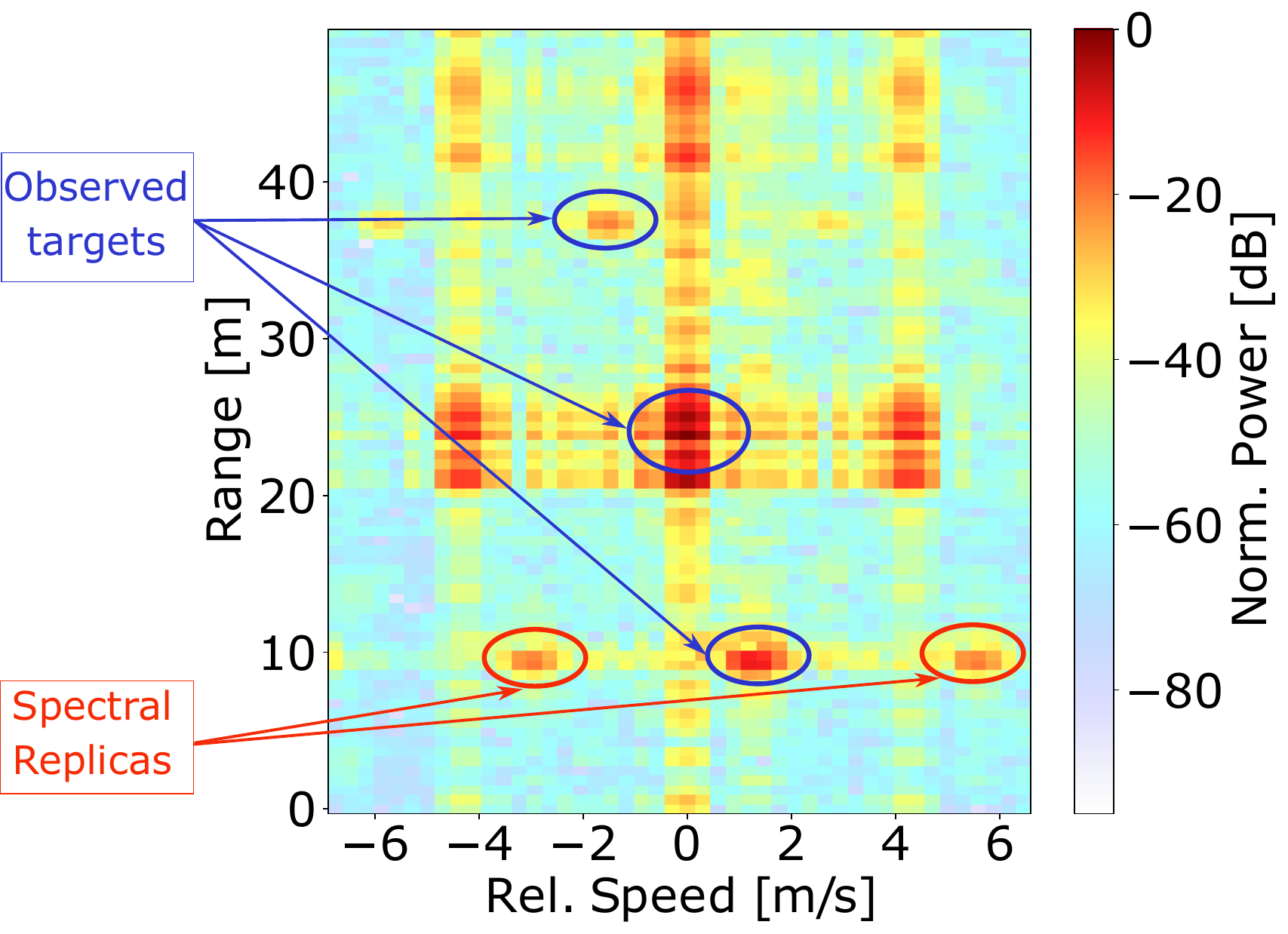}
    \caption{\small Periodogram after processing a full single CSI matrix. Target returns are highlighted in blue, spectral replicas due to empty UL symbols for a target at 10 m highlighted in red.  }
    \label{fig:TDD_patt_replicas}
\end{figure}

\begin{figure*}[t]
    \centering 
    \setlength{\abovecaptionskip}{5pt} 
    \setlength{\belowcaptionskip}{0cm} 
    \setlength{\intextsep}{1pt} 
    \begin{subfigure}[b]{0.327\textwidth}
        \includegraphics[width=\textwidth, trim={0.2cm 0.2cm 0.2cm 0.25cm},clip]{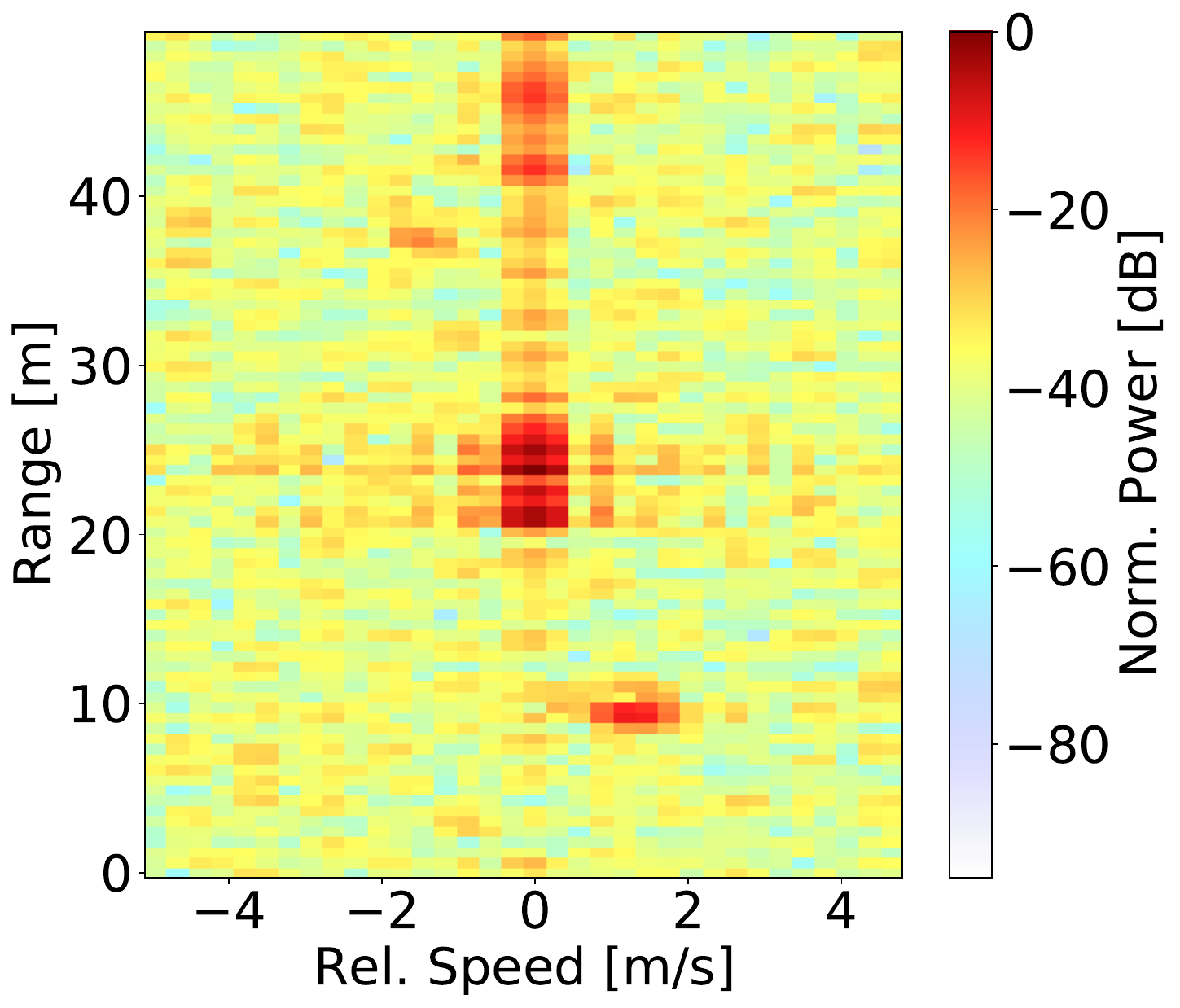}
	      \caption{Periodogram with single frame, decimated to 24 symbols ($J = 47$).} 
        \label{fig:tdd_per_single_dec}
    \end{subfigure}
    \hfill
    \begin{subfigure}[b]{0.327\textwidth}
        \includegraphics[width=\textwidth, trim={0.2cm 0.2cm 0.2cm 0.25cm},clip]{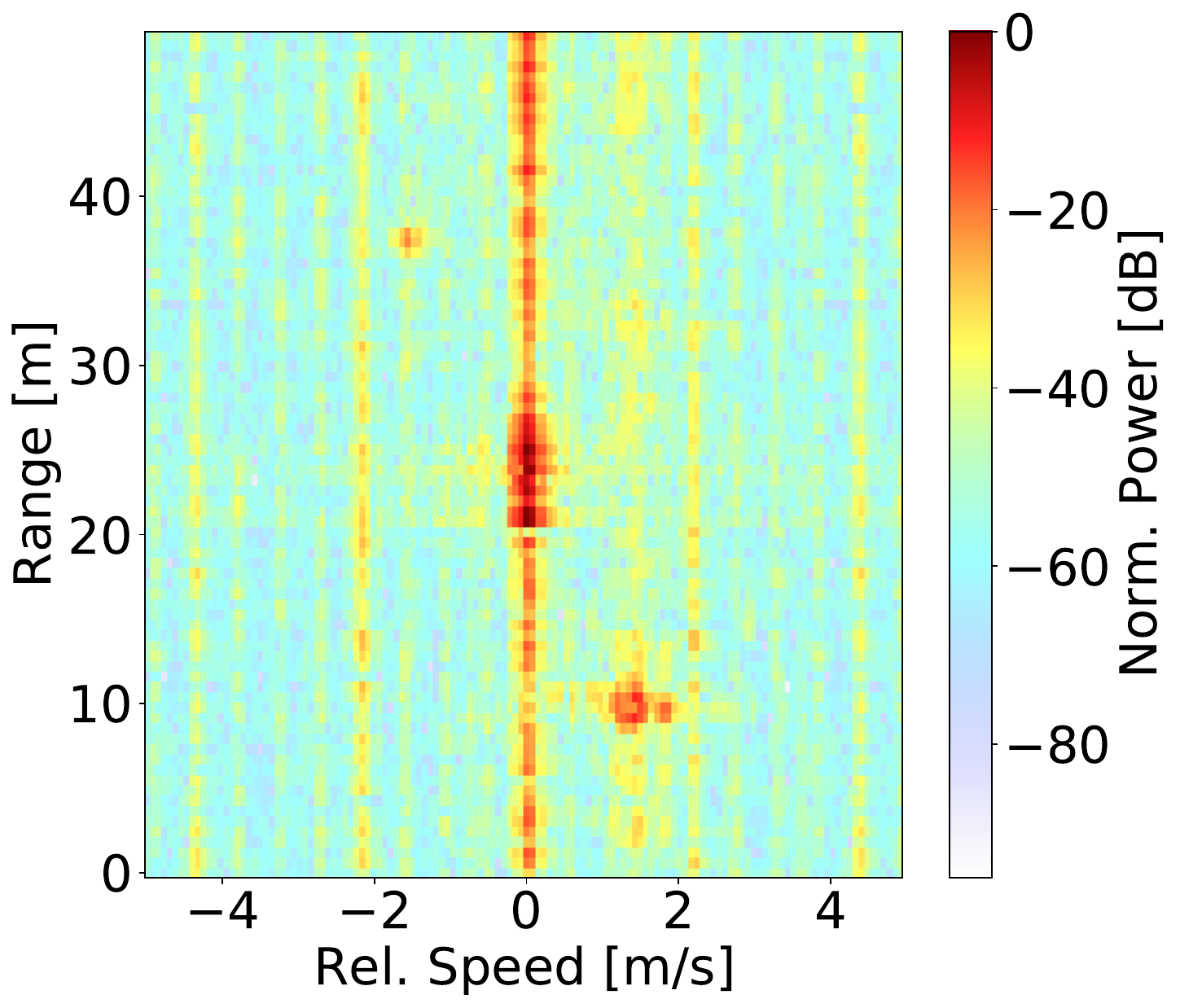}
	      \caption{Periodogram with 6 frames, decimated to 16 symbols per frame ($J = 70$) and concatenated.} 
        \label{fig:tdd_per_6_frames}
    \end{subfigure}
    \hfill
    \begin{subfigure}[b]{0.327\textwidth}
        \includegraphics[width=\textwidth, trim={0.2cm 0.2cm 0.2cm 0.25cm},clip]{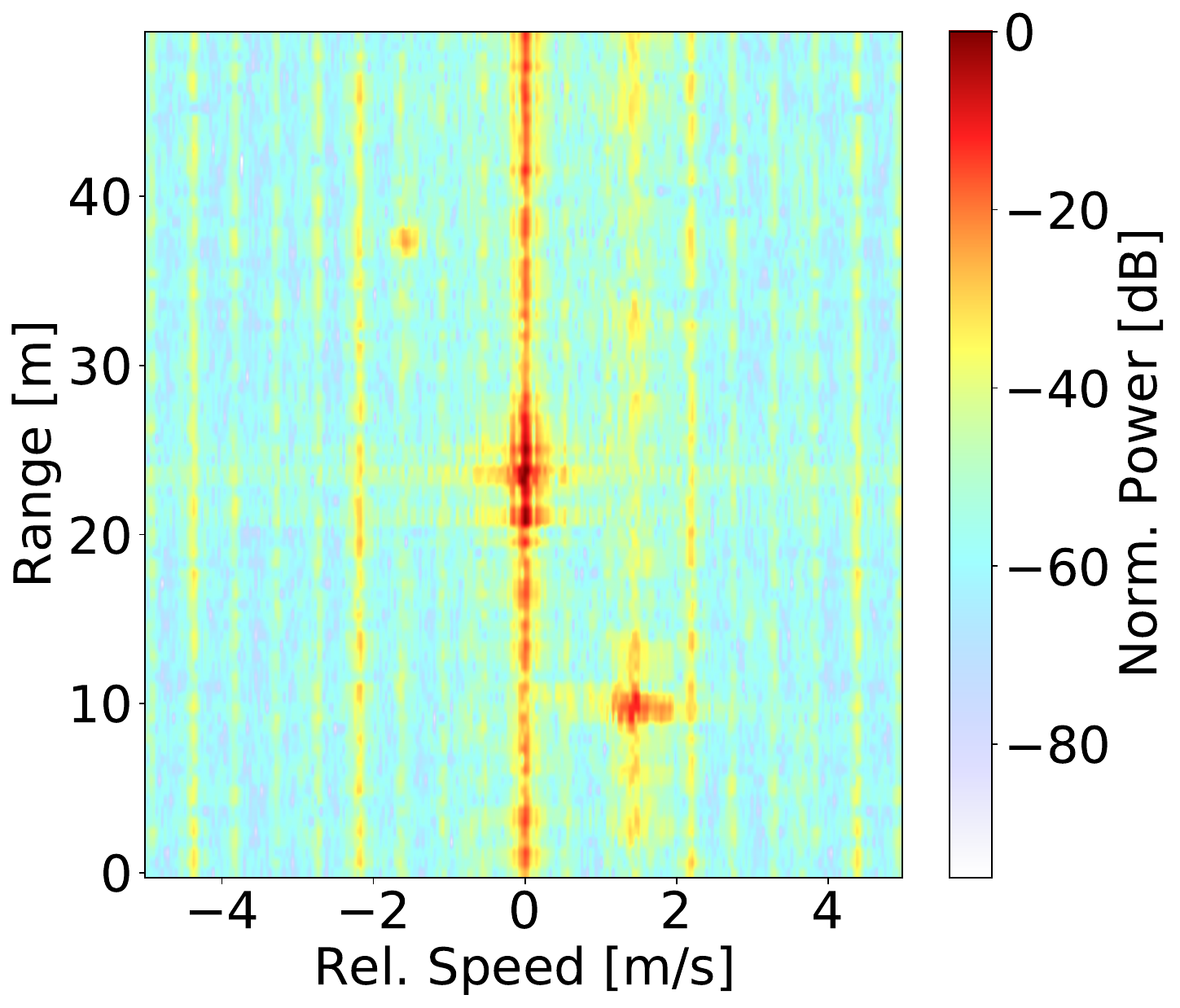}
	      \caption{Periodogram with 10 frames, decimated to 16 symbols per frame ($J = 70$) and concatenated.} 
        \label{fig:tdd_per_10_frames}
    \end{subfigure}
	\caption{\small Examples of possible decimation and combining approaches for periodogram processing. \vspace{-0.3cm}}
	\label{fig:tdd_strats}
\end{figure*}



After obtaining~$\mathbf H$, the range/Doppler periodogram is obtained by computing a DFT over the \gls{ofdm} symbols and an IDFT over the subcarriers~\cite{braun2014ofdm}
\begin{equation}
\begin{split}
S(n,m) &= \frac{1}{N'M'}\left|\sum_{k=0}^{N'} \Biggl(\sum_{l=0}^{M'} \mathbf{H}(k, l)e^{-j2\pi\frac{lm}{M'}}\Biggr)e^{j2\pi\frac{kn}{N'}} \right|^2, \\[-2ex]
\end{split}
\label{eq:periodogram}
\end{equation}
where $N' = 2^{\ceil{\log_2{N}}}$ and $M' = 2^{\ceil{\log_2{M}}}$ are the number of rows and columns of $\mathbf{H}$ after zero padding.
Note that sidelobes in the periodogram can be controlled by element-wise multiplication of $\mathbf H$ with a 2D window function. In our experiments, Chebyshev windowing was used.

In standard sensing operations, we process the whole \gls{csi} matrix $\mathbf H$ for each frame to maximize the processing gain and to achieve the speed resolution as in Eq.~\eqref{eq:speed_res}. However, since the \gls{gnb} transmits in a TDD pattern, \gls{ul} symbols do not contain useful information for radar processing and are discarded (set to zero) before computing the periodogram.
Using \gls{tdd}, each \gls{ul}/\gls{dl} pattern is repeated $ T_f/T_\text{TDD} = R$ times per frame. With the parameters from Section~\ref{sec:PoCScenario}, this amounts to $R=8$ \gls{tdd} patterns, as shown in Fig.~\ref{fig:CSI_with_TDD_patt}.
Setting the \gls{ul} parts to zero acts as an additional windowing effect on $\mathbf{H}$ and introduces spectral replicas in the speed domain, as shown in Fig.~\ref{fig:TDD_patt_replicas}.
This ``on/off" windowing is described by a discrete rectangle function, with a width of $M_{\text{DL}}$ symbols, convolved with a train of Dirac deltas in the time domain 
    \begin{align}
        w(t) &= \negthickspace \negthickspace\sum_{k = 0}^{M_{\text{DL}}-1} \negthickspace \negthickspace \delta(t-kT_s) \ast\negthickspace \negthickspace \sum_{k=-\infty}^\infty \negthickspace\delta\left( t - k T_{\text{TDD}} + \frac{M_{\text{DL}}T_s}{2} \right) \! .
        \label{eq:WindowingFunction}
    \end{align}
The resulting \textit{point spread function} is the Fourier transform of~\eqref{eq:WindowingFunction}, consisting of a Dirichlet Kernel 
multiplied by a train of Dirac deltas, that is convolved with the targets' contributions in the speed domain.  
As the \gls{tdd} pattern repeats $R$ times within the time aperture $MT_s$, the point spread function shows contributions in the speed domain spaced by 
\begin{equation}
\frac{c_0 R}{2 f_C M T_s} = R \cdot v_{\text{res}} = 4.4 \text{ m/s.}
\end{equation}
This is the source of the speed replicas that can be again observed to appear at $4.4$ m/s distance in Fig.~\ref{fig:TDD_patt_replicas}. 

\subsection{CSI Processing Techniques}

Processing only the first \gls{dl} sequence with $M_\text{DL}=104$ symbols would avoid replicas. However, per Eq.~\eqref{eq:speed_res}, this reduces the speed resolution to $5.88$~m/s, which is unacceptable for most use cases. 
Alternatively, the \gls{csi} matrix can be downsampled in time by selecting every \mbox{$J$-th} \gls{ofdm} symbol such that \gls{ul} parts are avoided. This approach is effective in removing the replicas but reduces the unambiguous speed by a factor of~$J$, as well as the processing gain due to the reduced number of processed symbols, leading to a lower \gls{snr}. 

\begin{table}[b]
    \caption{Periodogram SNR for different setups.} 
    \centering    
    \begin{tabu}{| c | c | c | c |}
    \hline
    Frames $K$ & Decimation $J$ & Symbols $M$ & \textbf{SNR [dB]} \\
    \Xhline{3\arrayrulewidth}
    \footnotesize 1 & 1 & 1120 & 60.91 \\ 
    \hline
    \footnotesize 1 & 47 & 24 & 45.16 \\ 
    \hline
    \footnotesize 6 & 70 & 96 & 52.35 \\ 
    \hline
    \footnotesize 10 & 70 & 160 & 54.66 \\ 
    \hline
    \end{tabu}
    \\
    \label{table:CSI_SNR}
\end{table}

\fakepar{Single frame processing}
By processing a single frame, it is possible to sample one symbol every $J$, thus with index~$m$, where $\mod(m,J) = 0$. For instance, $J = 47$ allows selecting 3 symbols for every \gls{dl} sequence and skipping the UL symbols completely. Fig.~\ref{fig:tdd_per_single_dec} shows an exemplary periodogram with this approach.
As the number of symbols is reduced from 1120 to 24, an \gls{snr} loss is visible compared to Fig.~\ref{fig:TDD_patt_replicas}. As presented in Tab.~\ref{table:CSI_SNR}, this amounts to a measured loss of 15.7~dB in the periodogram, in line with the expected $10\log_{10}(1120/24) = 16.7$~dB.
This loss can usually not be tolerated in \gls{nlos} scenarios, since the target return is typically weak due to the signal experiencing multiple reflections. With $J=47$, the unambiguous speed is reduced to $13.05$ m/s, which is still acceptable for most indoor use cases. 


\fakepar{Multiple frame processing}
To increase the processing gain, consecutive frames can be concatenated after decimating, increasing both the number of processed symbols and the time aperture of the acquisition. 
However, with \mbox{$J=47$}, only a single frame can be decimated before selecting \gls{ul} symbols.
By increasing the decimation to \mbox{$J=70$}, it is possible to select 2 symbols from every \gls{dl} section, with uniform rate over an indefinite number of frames without sampling \gls{ul} symbols. 
This allows to compute the periodogram from~$K$ consecutive frames concatenated together.
The resulting unambiguous speed of $8.76$ m/s is still enough for most indoor use cases.
As can be seen in Figs.~\ref{fig:tdd_per_6_frames} and Figs.~\ref{fig:tdd_per_10_frames}, this approach offers an increased \gls{snr} in the periodogram and a better speed resolution due to the increased time aperture. However, this comes at the cost of higher resource requirements and a lowered update rate. Compared to tracking and positioning tasks, in intrusion detection use cases a lowered update rate is less critical than a low \gls{snr}. 
However, it is possible, to retain a sufficiently large number of updates by letting consecutive observation windows overlap and setting an observation window stride $V<K$. 
Further, it should be noted that a large time aperture can lead to target migration effects, caused by the target's range and speed varying within the observation~window.

\section{NLOS experiments}\label{sec:nlos_experiments}
\subsection{NLOS Scenario}
\label{sec:meas_setup}

\begin{figure}[t]
    \centering
    \includegraphics[width=0.9\columnwidth, trim=0cm 0cm 0cm 0.2cm, clip]{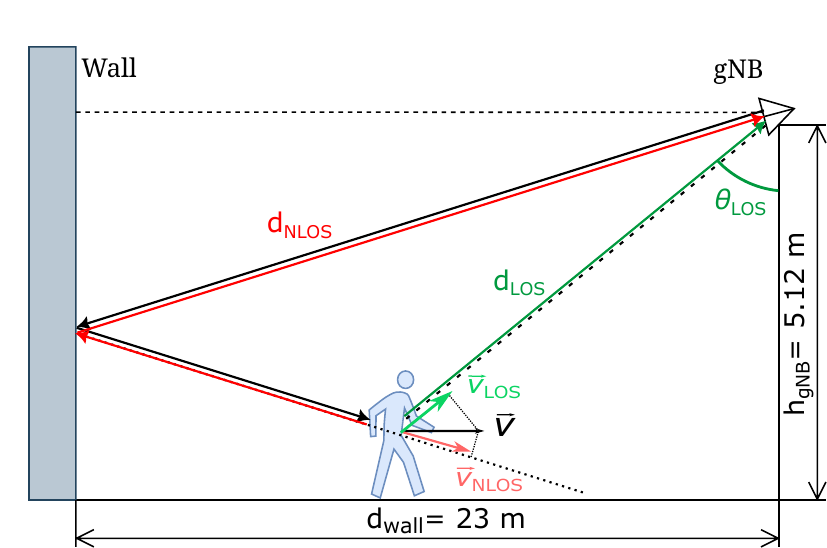}
    \caption{\small Geometry of the measurement scenario. The target moves back and forth in a straight line between the wall and the \gls{gnb}.}
    \label{fig:nlos-meas_scheme_lateral}
\end{figure}


The two \glspl{ru} of the \gls{isac} \gls{poc} are mounted at a height of $h_{\text{gNB}} = 5.12$~m and oriented towards a wide investigation area in the ARENA2036 industrial research campus in Stuttgart, Germany. The antenna pole is located at a horizontal distance of approximately $d_{\text{wall}}=23$~m from a warehouse rack (installed in front of a concrete wall) and a cargo door.
The measurement area is free of major obstructions that could be used to create a \gls{nlos} scenario under normal conditions. 
However, it is possible to emulate \gls{nlos} conditions by directing the signal over the target towards the wall and observing the reflected returns, as shown in Fig.~\ref{fig:nlos-meas_scheme_lateral}. Further, due to the large beamwidth, both \gls{los} and \gls{nlos} returns are present in our measurements. In our study, the \gls{los} component is used as a source of ground truth as a rough estimate of the target position. However, \gls{los} information has no bearing on the \gls{nlos} detection capabilities of our setup, but it is merely used as a simple remedy in the absence of accurate ground truth.

The scope of the experiments with the described setup was to determine the presence of the target via the \gls{nlos} return. 
At each update, given the geometry of the measurement described in Fig.~\ref{fig:nlos-meas_scheme_lateral}, the expected speed and range of the \gls{nlos} target return is estimated from the \gls{los} component as
\begin{align}
    \hat{v}_{\text{NLOS}} &\approx -v_{\text{LOS}} \; ,  \\
    \hat{d}_{\text{NLOS}} &= 2 d_{\text{wall}} - d_{\text{LOS}}\cdot \sin{\left(\theta_{\text{LOS}}\right)} \; ,
\end{align}
where $\theta_{\text{LOS}} = \arccos{\left(h_{\text{gNB}}/d_{\text{LOS}}\right)}$.
Peaks in the \gls{nlos} region are compared with the expected target position, allowing to determine whether they correspond to the return generated by the target or to interference.
If the \gls{nlos} range and speed match those estimated from ground truth, detection is positive.
The detection rate is computed as the ratio of frames in which the \gls{nlos} component was detected to the total number of frames in which the \gls{los} component was present.

\subsection{Radar Detection}
\label{sec:radar_detect}


After obtaining the periodogram with Eq.~\eqref{eq:periodogram}, it must be determined whether bins contain returns corresponding to targets of interest.
For this purpose, statistical tests are available in literature to decide whether a peak is due to noise only, or noise plus echos coming from a target.
The detection test used in this work belongs to the family of \gls{cfar} detectors \cite{Richards_Scheer_Holm_2010}.
Based on the observed interference level and a pre-defined probability of false alarm $p_\text{FA}$, standard \gls{cfar} defines a threshold for the whole periodogram as
\begin{equation}
    \eta_{\text{CFAR}} = - \sigma_N^2 \ln\left(1 - \left(1 - p_\text{FA}\right)^{\frac{1}{NM}}\right).
    \label{eq:cfar}
\end{equation}
In this work, we use a range-adjusted exponential threshold~\cite{Wagner_Feger_Stelzer_2017}, which is set by combining Eq.~\eqref{eq:cfar} with an $1/d^2$-shaped threshold to account for the propagation loss of target returns, which is proportional to the square of the distance to the RX.
The threshold for a bin corresponding to range $d$ is
\begin{align}
    \eta_\text{exp} = \eta_\text{CFAR} + \frac{\alpha}{d^2} \; ,
    \label{eq:ra_cfar_thresh}
\end{align}
where $\alpha$ is an adjustment factor for the exponential term. 
For this particular implementation, it is defined as the square root of the maximum return in the periodogram. The presence of a target is decided by comparing the signal level in every periodgram bin with this threshold.



\subsection{Clutter Removal}
\label{sec:clutter}
A known problem in radar is created by unwanted returns generated by objects in the environment, that are not of interest for the radar scope, and generally referred to as clutter.
In this work, clutter components are removed from the \gls{csi} matrix using subspace-based methods as described in~\cite{Henninger_CRAP_2023}.
The clutter removal matrices are obtained from calibration measurements with few acquisitions without targets. 
Data obtained from the clutter acquisition step can also be radar-processed, extracting information about the environment without targets, such as the distance of the wall/cargo gate in Fig.~\ref{fig:nlos-meas_scheme_lateral}.  


\subsection{NLOS Processing}
\label{subsec:NLOS_processing}
\begin{figure}[t]
    \centering
    \includegraphics[width=0.73\columnwidth, trim={0cm 0cm 0cm 1.4cm},clip]{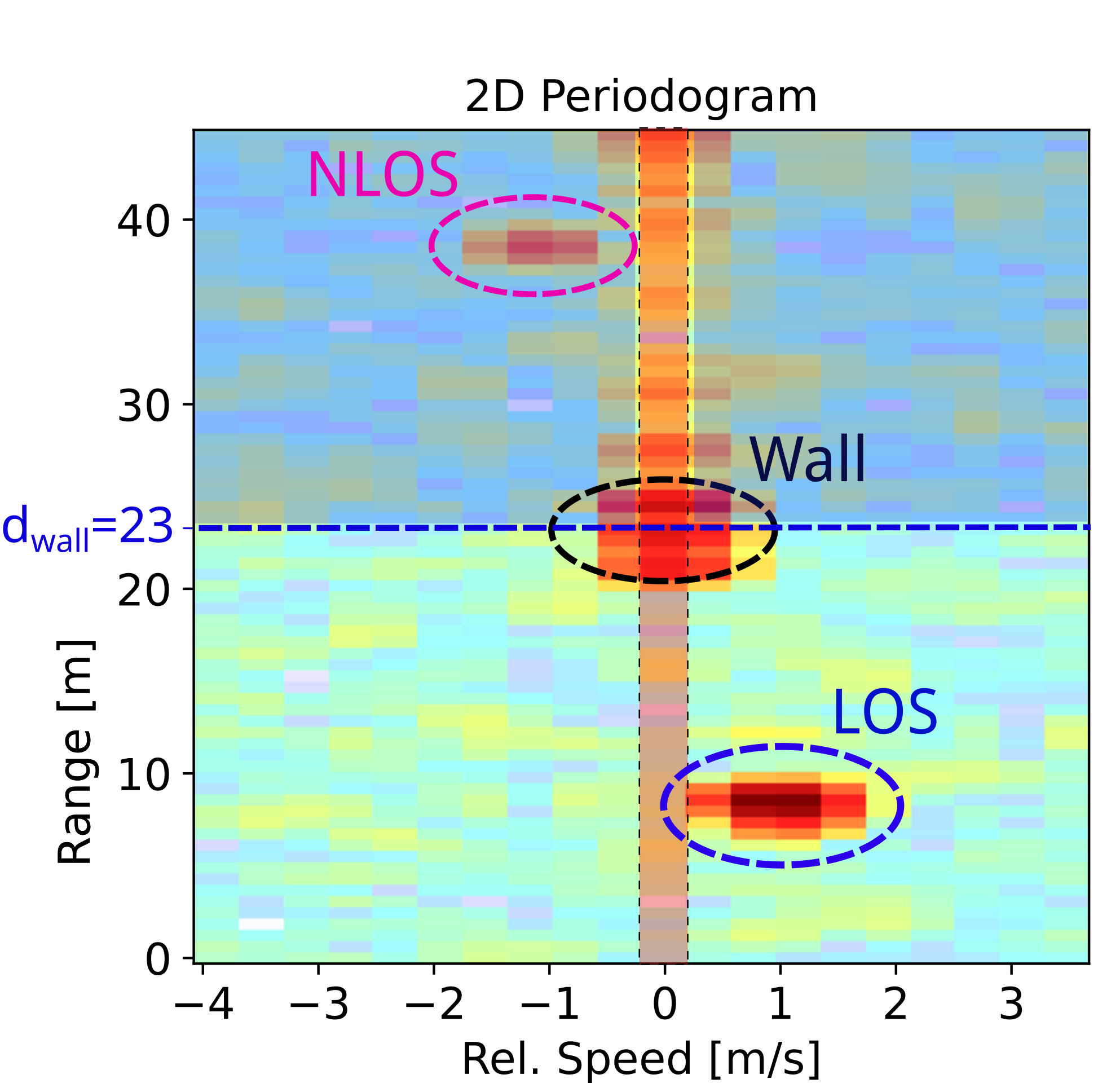}
    \caption{\small Exemplary periodogram from the scenario presented in Sec.~\ref{sec:meas_setup}.
    The static return at $d_{\text{wall}}$ is generated by the wall, and the joint \gls{los}-\gls{nlos} returns from the moving target are highlighted. The
    \gls{nlos} region is highlighted in blue, discarded static bins in red.}
    \label{fig:nlos-los-separation}
\end{figure}

\Gls{los} sensing capabilities of our system have been demonstrated in \cite{wild2023integrated}. The focus is now on whether it is possible to monitor areas in \gls{nlos} to enable intrusion detection services with \gls{isac}.
For this, the returns in the periodogram generated by targets in \gls{nlos} conditions must be determined.
To ensure \gls{nlos} coverage, the signal must be directed towards a large obstacle with a sufficient 
radar cross-section before being reflected towards the target, similar to Fig.~\ref{fig:nlos-meas_scheme_lateral}.
Any return with a range greater than the wall can then be considered a \gls{nlos} return.
However, distinguishing static targets from clutter can be challenging, as static clutter is generated by other multipath reflections. 
For \gls{nlos} intrusion detection, it is sufficient to detect an intruder only when it moves. 
Therefore, bins of the periodogram corresponding to null speed are discarded for \gls{nlos} processing.
After extracting the position of the wall during calibration, the periodograms processed during online sensing operations are divided into two regions, corresponding to \gls{los} and \gls{nlos} conditions, respectively.
Fig.~\ref{fig:nlos-los-separation} highlights the separation for a sample periodogram, where the return from the moving target at ca. 40 m is considered \gls{nlos}.

\section{Measurement Results}
\label{sec:meas_result}


We tested the \gls{nlos} detection capabilities on human targets, moving in the emulated \gls{nlos} environment between wall and \gls{gnb} (Fig.~\ref{fig:nlos-meas_scheme_lateral}).
The results were obtained by concatenating multiple frames after decimating the \gls{csi} as presented in Sec.~\ref{sec:csi_strategies}.
After computing the periodogram, targets were detected using the range-adjusted exponential threshold from Eq.~\eqref{eq:ra_cfar_thresh}.

Detection rate was estimated for the 10 strongest \gls{nlos} returns, comparing them with the \gls{los} components.
%
%
With the focus on intrusion detection, it is sufficient to detect the presence of the intruder once and raise an alarm accordingly.
A possible approach for discriminating the presence of actual targets from false alarms is to verify that \gls{nlos} peaks are continuously detected within a window of consecutive updates.
In our measurements, an overall detection rate of up to 67\% was obtained with $K=8$ frames and stride $V=2$ frames.

\begin{figure}[t]
    \setlength{\abovecaptionskip}{3pt} 
    \setlength{\belowcaptionskip}{3pt} 
    \setlength{\intextsep}{4pt} 
    \centering
    \includegraphics[width=0.85\columnwidth, height = 5.2cm, trim=0.4cm 0.4cm 0.35cm 1.05cm, clip]{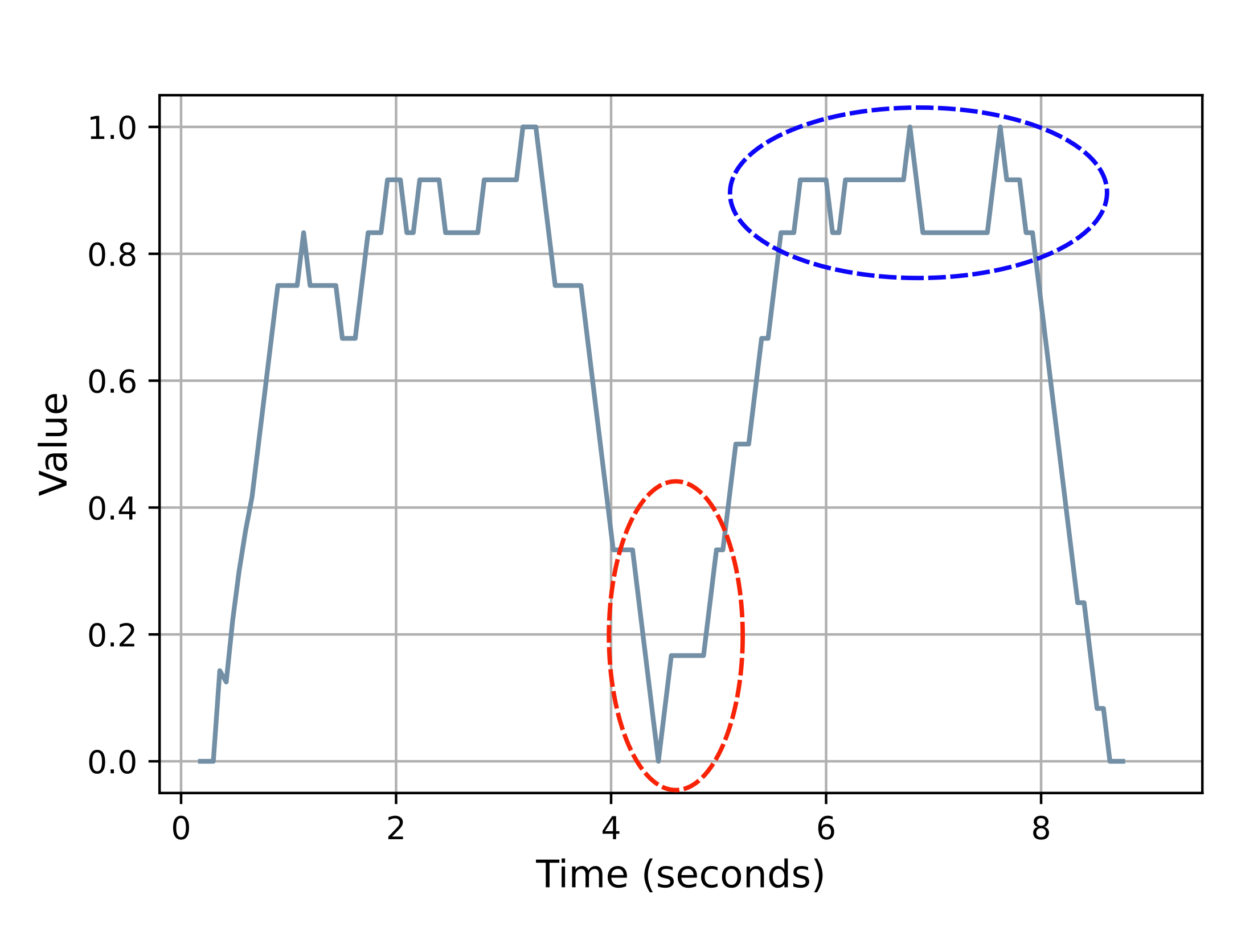}
    \caption{Moving average of NLOS detection rate over a 0.5 s time window, using exponential threshold and clutter removal. Periodograms were generated using CSI matrices from $K=8$ frames.
    }
    \label{fig:detect_mov_avg}
\end{figure}

The sustained presence of the \gls{nlos} return over an observation window can be evaluated by computing the moving average of the detection rate over time. 
Fig.~\ref{fig:detect_mov_avg} depicts the detection rate over a 0.5~s time window across the measured scenario.
The calculation of the detection rate includes the parts of the measurements where the target stopped and changed direction (highlighted in red in Fig.~\ref{fig:detect_mov_avg}).
In these instances, no \gls{nlos} component is detected, as the bins of the periodogram corresponding to zero speed are discarded as described in~\ref{subsec:NLOS_processing}.
It can be observed that the \gls{nlos} component is detected in a sustained way when the target is moving, allowing reliable detection with further processing.

\section{Conclusion}

In this work, we investigated \gls{nlos} sensing for \gls{isac}. Based on measurements with a \gls{mmw} \gls{poc} in a factory-like environment, we showed that target detection in \gls{nlos} is generally possible, enabling promising use cases like intrusion detection. To achieve that, we designed a \gls{csi} processing and detection routine that allows coping with the specifics of communications systems, while attaining the necessary \gls{snr} and resolution to detect moving objects in \gls{nlos}.

In the future, we will refine the detection routine, integrating also tracking techniques to further increase the reliability of the approach. Moreover, we plan to validate our method with additional measurements in other scenarios.


\section*{Acknowledgments}
The authors would like to thank Artjom Grudnitsky for the fruitful discussions during the development of this work and for his help in taking the measurements.

This work was developed within the KOMSENS-6G project, partly funded by the German Ministry of Education and Research under grant 16KISK112K.

\bibliographystyle{IEEEtran}
\bibliography{nlos.bib}

\begin{thebibliography}{10}
\providecommand{\url}[1]{#1}
\csname url@samestyle\endcsname
\providecommand{\newblock}{\relax}
\providecommand{\bibinfo}[2]{#2}
\providecommand{\BIBentrySTDinterwordspacing}{\spaceskip=0pt\relax}
\providecommand{\BIBentryALTinterwordstretchfactor}{4}
\providecommand{\BIBentryALTinterwordspacing}{\spaceskip=\fontdimen2\font plus
\BIBentryALTinterwordstretchfactor\fontdimen3\font minus \fontdimen4\font\relax}
\providecommand{\BIBforeignlanguage}[2]{{%
\expandafter\ifx\csname l@#1\endcsname\relax
\typeout{** WARNING: IEEEtran.bst: No hyphenation pattern has been}%
\typeout{** loaded for the language `#1'. Using the pattern for}%
\typeout{** the default language instead.}%
\else
\language=\csname l@#1\endcsname
\fi
#2}}
\providecommand{\BIBdecl}{\relax}
\BIBdecl

\bibitem{Mandelli_Henninger_Bauhofer_Wild_2023}
S.~Mandelli, M.~Henninger, M.~Bauhofer, and T.~Wild, ``{Survey on Integrated Sensing and Communication Performance Modeling and Use Cases Feasibility},'' in \emph{2023 2nd Int. Conf. on 6G Netw.}, Oct. 2023.

\bibitem{Wang_isac_survey_2022}
J.~Wang, N.~Varshney, C.~Gentile, S.~Blandino, J.~Chuang, and N.~Golmie, ``\BIBforeignlanguage{en}{{Integrated Sensing and Communication: Enabling Techniques, Applications, Tools and Data Sets, Standardization, and Future Directions}},'' \emph{\BIBforeignlanguage{en}{IEEE Internet of Things J.}}, vol.~9, no.~23, p. 23416–23440, Dec. 2022.

\bibitem{Li_Ge_Wang_BreathingNLOS_2022}
G.~Li, Y.~Ge, Y.~Wang, Q.~Chen, and G.~Wang, ``\BIBforeignlanguage{en}{{Detection of Human Breathing in Non-Line-of-Sight Region by Using mmWave FMCW Radar}},'' \emph{\BIBforeignlanguage{en}{IEEE Trans. on Instrum. and Meas.}}, vol.~71, p. 1–11, Sep. 2022.

\bibitem{Gustafsson_Doppler_urban_2016}
M.~Gustafsson, A.~Andersson, T.~Johansson, S.~Nilsson, A.~Sume, and A.~Örbom, ``{Extraction of Human Micro-Doppler Signature in an Urban Environment Using a “Sensing-Behind-the-Corner” Radar},'' \emph{IEEE Geosci. and Remote Sens. Lett.}, vol.~13, no.~2, p. 187–191, Feb. 2016.

\bibitem{Solomitckii_nlos_uwb}
D.~Solomitckii, C.~B. Barneto, M.~Turunen, M.~Allén, Y.~Koucheryavy, and M.~Valkama, ``{Millimeter-Wave Automotive Radar Scheme With Passive Reflector for Blind Corner Conditions},'' in \emph{2020 14th Eur. Conf. on Antennas and Propag.}, 2020, pp. 1--5.

\bibitem{pegoraro2024jump}
J.~Pegoraro \emph{et~al.}, ``{JUMP: Joint communication and sensing with Unsynchronized transceivers Made Practical},'' \emph{IEEE Transactions on Wireless Communications}, Feb. 2024, early access, \mbox{doi}:\url{10.1109/TWC.2024.3365853}.

\bibitem{wild2023integrated}
T.~Wild, A.~Grudnitsky, S.~Mandelli, M.~Henninger, J.~Guan, and F.~Schaich, ``{6G Integrated Sensing and Communication: From Vision to Realization},'' in \emph{2023 20th Eur. Radar Conf.}, Sep. 2023, pp. 355--358.

\bibitem{3gpp_38331}
3GPP, ``{NR; Radio Resource Control (RRC); Protocol specification},'' Technical Specification (TS) 38.331, 2023, version 17.5.0.

\bibitem{3gpp_38211}
------, ``{NR; Physical channels and modulation},'' Technical Specification (TS) 38.211, 2023, version 17.4.0.

\bibitem{de2021joint}
L.~Giroto~de Oliveira, B.~Nuss, M.~B. Alabd, A.~Diewald, M.~Pauli, and T.~Zwick, ``{Joint radar-communication systems: Modulation schemes and system design},'' \emph{IEEE Trans. on Microwave Theory and Techn.}, vol.~70, no.~3, pp. 1521--1551, Mar. 2022.

\bibitem{braun2014ofdm}
K.~M. Braun, ``{OFDM Radar Algorithms in Mobile Communication Networks},'' Ph.D. dissertation, Karlsruher Institut f{\"u}r Technologie, 2014.

\bibitem{Richards_Scheer_Holm_2010}
M.~A. Richards, J.~A. Scheer, and W.~A. Holm, \emph{{Principles of Modern Radar: Basic Principles}}.\hskip 1em plus 0.5em minus 0.4em\relax Raleigh, NC, USA: SciTech Pub., 2010.

\bibitem{Wagner_Feger_Stelzer_2017}
T.~Wagner, R.~Feger, and A.~Stelzer, ``\BIBforeignlanguage{en}{{Radar Signal Processing for Jointly Estimating Tracks and Micro-Doppler Signatures}},'' \emph{\BIBforeignlanguage{en}{IEEE Access}}, vol.~5, p. 1220–1238, Feb. 2017.

\bibitem{Henninger_CRAP_2023}
M.~Henninger, S.~Mandelli, A.~Grudnitsky, T.~Wild, and S.~ten Brink, ``{CRAP: Clutter Removal with Acquisitions Under Phase Noise},'' in \emph{2023 2nd Int. Conf. on 6G Networking}, Oct. 2023.

\end{thebibliography}
\balance

\end{document}